# On the Nuclear Mechanisms Underlying the Heat Production by the E-Cat

**Norman D. Cook**[1] **and Andrea Rossi**[2]

1. Department of Informatics, Kansai University, Osaka, 1095-569, Japan
2. Leonardo Corporation, Miami Beach, Florida, 33139, USA

We discuss the isotopic abundances found in the E-Cat reactor with regard to the nuclear mechanisms responsible for excess heat. We argue that a major source of energy is a reaction between the first excited-state of Li-7 and a proton, followed by the breakdown of Be-8 into two alphas with high kinetic energy, but without gamma radiation. The unusual property of the Li-7 isotope that allows this reaction is similar to the property that underlies the Mossbauer effect: the presence of unusually low-lying excited states in stable, odd-Z and/or odd-N nuclei. We use the lattice version of the independent-particle model (IPM) of nuclear theory to show how the geometrical structure of isotopes indicate nuclear reactions that are not predicted in the conventional version of the IPM. Finally, we speculate on similar mechanisms that may be involved in other low-energy nuclear reactions (LENR).

## 1 Introduction

The checkered history of low-energy nuclear reaction (LENR) research remains highly controversial. It includes disputed claims of both experimental successes and failures in both Nickel and Palladium systems. Reported results and theoretical models are far too diverse to allow definitive conclusions to be drawn, but Storms [1, 2] has summarized the overwhelming consensus that nuclear effects have been obtained in experimental set-ups where conventional theory predicts the total absence of nuclear involvement. While further empirical work remains a high priority, a remaining theoretical task is to demonstrate how the published data on heat production and isotopic transmutations are consistent with the major themes of nuclear physics, as established over the past century.

In the latest empirical test of Andrea Rossi's invention, known as the E-Cat, significant excess heat (a ratio of output/input energy in excess of 3.0) over the course of one month was found [3]. For technological exploitation, it may be sufficient to mimic the materials and protocols that have made that possible (e.g., Parkhomov [4]), but the huge diversity of conditions that have been reported in the "cold fusion" literature for 26 years suggest that there may exist general LENR mechanisms that have not yet been identified. Although progress has been made in defining the solid-state, chemical and electromagnetic field properties of the nuclear active environment (NAE), the specifically **nuclear** aspects of the NAE have not generally been addressed. Here, we argue that femtometer-level LENR can occur in isotopes with low-lying excited-states, provided that an appropriate, Angstrom-level molecular environment has been created.

In the present study, we focus on recent findings of nuclear transmutations concerning Lithium isotopes [3] in light of the lattice version [5] of the independent-particle model (IPM) of the nucleus. Specifically, after brief review of the well-established IPM, we consider details of the substructure of the $^7_3Li_4$ and $^{8}_{4}Be_{4}$ isotopes that allow for the generation of alpha particles at kinetic energies well beyond what could be produced solely through chemical reactions.

# 2 Methods

## 2.1 Theory: The Independent-Particle Structure of Nuclei

For more than six decades, it has been known that many nuclear properties can be described in terms of the simple summation of the properties of the constituent protons and neutrons. In the 1930s, this theoretical perspective was rejected by Niels Bohr, who favored a "collective" view of nuclei, but the shell model assumption of spin-orbit coupling in the early 1950s proved to be a major theoretical success that established the "independent-particle" approach as the central paradigm of nuclear structure theory.

Most importantly, the IPM description of nuclear states allowed for a coherent explanation of experimentally observed spins and parities ($J\pi$) (and, more approximately, magnetic moments, $\mu$) as the summation of the $j\pi$ and $\mu$ of any unpaired protons and neutrons. Subsequently, the ground-state spin and parity of more than 2800 relatively-stable nuclear isotopes and, most impressively, the nearly half-million excited-states of those isotopes, as tabulated in the *Firestone Table of Isotopes* (1996) [6], have been classified in the IPM. Arguably, it is this undisputed success of the IPM that has led many nuclear physicists to conclude that LENR phenomena are unlikely to be real, insofar as they are not consistent with the established principles of nuclear theory. As discussed below, we have found that the theoretical framework provided by the IPM is, on the contrary, essential for explaining the transmutations reported to occur in the E-Cat.

The early mathematical development of the IPM was undertaken by Eugene Wigner [7] in the 1930s, but the IPM did not become the dominant model of nuclear structure until the early 1950s, with the emergence of the shell model [8]. In fact, Wigner and the inventors of the shell model shared the Nobel Prize in Physics in 1963, and their combined insights gave nuclear structure theory a coherent quantum mechanical basis. The bold assumption of the shell model was that there occurs a coupling between quantum numbers, *l* and *s*, to produce an *observable* total angular momentum, *j* ($=l\pm s$). Inherent to that assumption, however, was the highly *unrealistic* notion that "point" nucleons "orbit" freely in the nuclear interior and do not interact with other nucleons (in first approximation) that as orbiting within the nuclear potential well. Similar assumptions had also been made in the still earlier Fermi gas model of the nucleus, but were eventually rejected because of the theoretical successes of the liquid-drop model (LDM) concerning nuclear binding energies, radii, fission phenomena, etc. The LDM, in turn, was based upon *realistic* assumptions about the nuclear interior: electrostatic and magnetic RMS radii of protons and neutrons of about 0.85 fm [9], a nuclear core density of 0.17 nucleons/fm$^3$ (implying a nearest-neighbor internucleon distance of only 2.0 fm) and the ***non-orbiting*** of nucleons – all of which argued strongly ***against*** a diffuse nuclear "gas" and ***for*** a dense nuclear "liquid".

The inherent contradictions between the gaseous-phase IPM and the liquid-phase LDM are of course summarized in most nuclear textbooks, but an interesting blend of those two competing models was first developed in the 1970s in the form of a lattice model of nuclear structure (the history of which is discussed in ref. [5]). The lattice model (a "frozen liquid-drop") has most of the properties of the traditional LDM, but, when nuclei are built around a central tetrahedron of four nucleons, the lattice shows the remarkable property of reproducing the correct sequence and occupancy of all of the *n*-shells of the shell model as triaxially-symmetrical (spherical) lattice structures. Moreover, the *j*-subshells within the *n*-shells of the shell model emerge as cylindrical structures and the *m*-subshells arise as conical substructures – all in the same sequence and with the same occupancy of protons and neutrons as known from the conventional IPM.

Although various aspects of the mathematical identity between the shell and lattice models have frequently been published in the physics literature, the lattice model itself has been dismissed as a "lucky" reproduction of the symmetries of the shell model and has had little impact on nuclear theorizing, in general. The fact remains, however, that the lattice and gaseous-phase versions of the IPM reproduce the same patterns of *observable* spin and parity ($J\pi$) values based upon very different assumptions concerning the "point" or "space-occupying" structure of the nucleons themselves. Here, we consider the lattice IPM to be a realistic alternative to the gaseous-phase IPM, and elaborate on its implications in relation to LENR phenomena.

The quantal properties in the lattice model are defined in Eqs. (1-6), and are illustrated in Figure 1. Related theoretical arguments have been published since the 1970s, and full details are available online [5].

$$n = (|x| + |y| + |z| - 3) / 2 \qquad \text{(Eq. 1)}$$
$$l = (|x| + |y|) / 2 \qquad \text{(Eq. 2)}$$
$$j = (|x| + |y| -1) / 2 \qquad \text{(Eq. 3)}$$
$$m = |y| *(-1)^{((x-1)/2)}/ 2 \qquad \text{(Eq. 4)}$$
$$s = (-1)^{((x-1)/2)}/ 2 \qquad \text{(Eq. 5)}$$
$$i = (-1)^{((z-1)/2)} \qquad \text{(Eq. 6)}$$

The significance of the "quantal geometry" (Eqs. 1-6) (Figure 1) can be simply stated: every unique grid site in the lattice corresponds to a unique set of nucleon quantum numbers, the sum of which is identical to that produced in the conventional IPM. Conversely, knowing the quantum characteristics of individual nucleons, their positions (Cartesian coordinates) in the lattice can be calculated, as shown in Eqs. (7-9).

$$x = |2m|(-1)^{(m-1/2)} \qquad \text{(Eq. 7)}$$
$$y = (2j+1-|x|)(-1)^{(i/2+j+m+1/2)} \qquad \text{(Eq. 8)}$$
$$z = (2n-3+|x|-|y|)(-1)^{(i/2+n+j+1)} \qquad \text{(Eq. 9)}$$

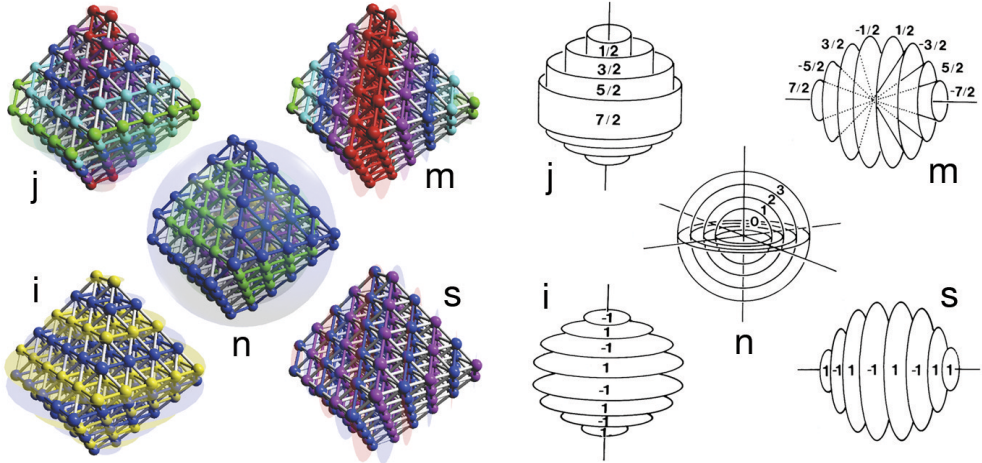

Figure 1: The geometry of nuclear quantum numbers in the lattice representation of the IPM.

A simple example of the identity between IPM quantal features and lattice symmetries is illustrated in Figure 2 for the ground-state of $^{15}_{7}N_8$. On the left is shown the build-up of protons and neutrons in a conventional tabulation of IPM states in relation to the quantum numbers. On the right is shown the corresponding lattice structure for those 15 nucleons. Note that the geometrical configuration of neutrons (blue) and protons (yellow) is given explicitly by the lattice definitions of the quantum numbers. In other words, the configuration of nucleons in the lattice IPM is determined by the quantum characteristics of the given isotope's nucleons.

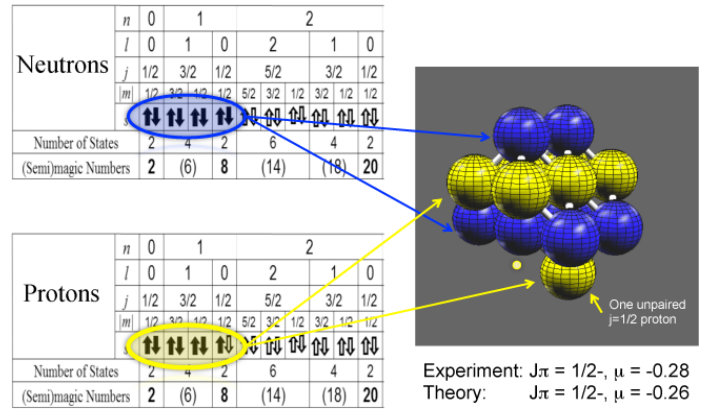

Figure 2: The IPM quantal states of the 8 neutrons and 7 protons of $^{15}_{7}N_8$ (the filled arrows on the left) and their lattice positions (right), as determined from Eqs. (1-6). The unpaired proton is responsible for the spin/parity and magnetic moment predictions of the IPM; the lone unfilled proton site (-1, -1, -3) in the second *n*-shell is shown as a dot.

In the same way that there is a precise identity between IPM states and lattice configurations for all ground-state nuclei, excited-states have corresponding lattice structures whose spins/parities are identical to those measured experimentally. For example, the nine lowest-lying states of $^{15}_{7}N_8$ are shown in Figure 3.

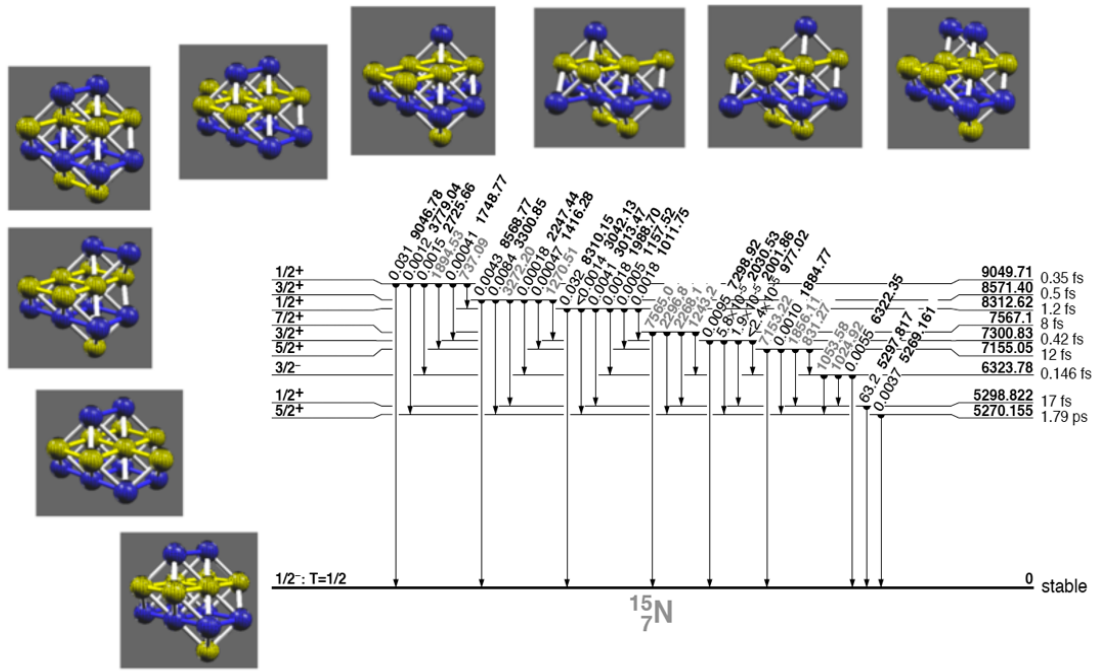

Figure 3: The low-lying excited states of $^{15}_{7}N_8$, and their corresponding lattice structures. Every lattice structure is a unique set of proton and neutron sites, whose *j*π values sum to the measured *J*π values that are known from experiment and shown in the level diagram.

## 2.2 Experiment: The New Transmutation Data

In the recent Lugano report on the E-Cat [3], two types of nuclear transmutation were noted (Table 1). Similar isotopic changes have also been reported by Parkhomov [4], lending credence to the earlier report, but neither experimental study discussed possible theoretical nuclear mechanisms. The first type of transmutation was a strong decrease in $^7_3Li_4$ relative to the only other stable isotope of Lithium, $^6_3Li_3$. The second was a strong relative increase in one Nickel isotope, $^{62}_{28}Ni_{34}$, and large relative decreases in $^{58}_{28}Ni_{30}$ and $^{60}_{28}Ni_{32}$, accompanied by small, but significant decreases in $^{61}_{28}Ni_{33}$ and $^{64}_{28}Ni_{36}$ (Table 1). These effects need to be explained within the framework of conventional nuclear theory.

The dilemma that theorists face is that both excess heat production and altered isotopic ratios are strongly suggestive of nuclear involvement, but conventional theory alone provides no clue on how these nuclear reactions could arise. While this theoretical stalemate remains unresolved, however, we demonstrate below how specific isotopic structures in the lattice IPM could in principle lead to the strong depletion of $^7_3Li_4$, while implying the generation of alpha particles – provided only that energetic justification for such effects can be found from basic theory.

**Table 1:** Transmutations at the Onset and Conclusion of the E-Cat Test [1]

| Isotope | Natural Abundance | Abundance at Onset | Abundance at Conclusion |
|---|---|---|---|
| $^6_3Li_3$ | 7.5% | 8.6% | 92.1% |
| $^7_3Li_4$ | 92.5% | 91.4% | 7.9% |
| $^{58}_{28}Ni_{30}$ | 68.077% | 67.0% | 0.8% |
| $^{60}_{28}Ni_{32}$ | 26.223% | 26.3% | 0.5% |
| $^{61}_{28}Ni_{33}$ | 1.140% | 1.9% | 0.0% |
| $^{62}_{28}Ni_{34}$ | 3.634% | 3.9% | 98.7% |
| $^{64}_{28}Ni_{36}$ | 0.926% | 1.0% | 0.0% |

In theory, the changes in Lithium isotopes could be a consequence of three distinct mechanisms: (i) *de novo* creation of $^6_3Li_3$ (leading to relative increases in this isotope) (ii) the transmutation of $^6_3Li_3$ and/or $^7_3Li_4$ by the addition/removal of one neutron (leading to relative increases and decreases, respectively), or (iii) *de novo* destruction of $^7_3Li_4$ (leading to relative decreases in this isotope).

*De novo* creation of $^6_3Li_3$ (i) is the most problematical, because it implies the sequential accretion of protons and neutrons; low energy mechanisms of that type are unknown. Similarly, the transmutation of $^6_3Li_3$ into $^7_3Li_4$ or vice versa (ii) requires the accretion or depletion of neutrons in an experimental set-up where free neutrons have not been detected; nuclear mechanisms of that type are also unknown.

*De novo* destruction of $^7_3Li_4$ (iii), in contrast, is theoretically plausible, insofar as the accretion of one proton would transmute $^7_3Li_4$ into $^8_4Be_4$, which could then decay to two alphas with the release of significant kinetic energy, leading to a relative decrease in $^7_3Li_4$:

$$^7_3Li_4 + p \rightarrow \, ^8_4Be_4 \rightarrow 2\alpha \,(17.26 \text{ MeV}) \quad \text{(Eq. 10)}$$

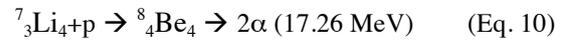

The question of energetic mechanisms aside, the depletion of $^7_3Li_4$ through the accretion of one proton is a theoretical possibility insofar as it does not imply gamma radiation. That is to say, the decay of $^8_4Be_4$ to two alpha particles is known to be gamma-free. Provided that the initial approach of a proton to the Lithium isotope can be energetically justified, the formation of $^8_4Be_4$ and the subsequent generation of energetic alphas would therefore not be problematical. Clearly, an abundance of such reactions would lead to four observable effects: (i) absolute decreases in $^7_3Li_4$ with (ii) relative increases in $^6_3Li_3$, together with (iii) the generation of alpha particles, and (iv) the production of significant kinetic energies, as the alphas are repelled from one another.

## 2.3 Using the Lattice IPM to Explain LENR

While the IPM accurately specifies the properties of excited-states (as was illustrated for the level diagram of $^{15}_{7}N_{8}$, Figure 3), the conventional Fermi-gas-like perspective on nuclear structure explicitly denies the possibility of nuclear substructure (beyond that implied by deformations of the nuclear potential well).

In contrast, the lattice representation of the IPM makes precisely the same predictions concerning the quantal properties of nuclei (Eqs. 1-6), but the lattice structures can also be used to specify the "stereochemical" structure of nuclei. In other words, because there are specific, often unique, lattice structures corresponding to each and every ground- and excited-state, the lattice version of the IPM provides candidate structures that are involved in various nuclear reactions. If the high-temperature, high-pressure conditions within the E-Cat provide sufficient energy to allow Hydrogen nuclei to overcome the Coulomb barrier and to approach Lithium nuclei, then the Lithium nucleus itself may be promoted to a low-lying excited state. An interaction between Hydrogen and Lithium nuclei within appropriate solid-state environments could then be accompanied by certain types of LENR that depend principally on the detailed substructure of the Lithium isotope.

# 3 Results

## 3.1 Lithium Transmutations

The two stable Lithium isotopes, $^{6}_{3}Li_{3}$ and $^{7}_{3}Li_{4}$, are well characterized in the IPM in terms of their constituent particles (Table 2). Given the IPM properties of the nucleons around the $^{4}_{2}He_{2}$ core, their fine-structure in the lattice IPM is unambiguous.

Table 2: The substructure of the Lithium isotopes. Generally, the IPM description of isotopes gives properties close to those measured experimentally (spin/parities matching empirical values and magnetic moments within 20% of empirical values).

| Isotope | Binding Energy | Spin/Parity | Magnetic Moment | RMS Radius |
|---|---|---|---|---|
| **Li-6** | 31.994 MeV | **1+** | +0.822 | 2.589 fm |
| [ He-4 | 28.296 MeV | 0+ | 0.000 | 1.676 fm ] |
| [ p |  | 1/2+ | +2.793 | 0.865 fm ] |
| [ n |  | 1/2+ | -1.914 | 0.873 fm ] |
| Theory: |  | **1+** | +0.889 | 2.545 fm |
|  |  |  |  |  |
| **Li-7** | 39.244 MeV | **3/2-** | +3.256 | 2.444 fm |
| [ He-4 | 28.296 MeV | 0+ | 0.000 | 1.676 fm ] |
| [ p |  | 3/2- | +3.793 (Schmidt) | 0.865 fm ] |
| [ n |  | 1/2+ | -1.914 | 0.873 fm ] |
| [ n |  | 1/2+ | +1.914 | 0.873 fm ] |
| Theory: |  | **3/2-** | +3.793 | 2.550 fm |

That is, both the spin/parity and the magnetic moments of these nuclei can be understood simply as the summation of the properties of a $^{4}_{2}He_{2}$ core plus a few additional nucleons. For $^{6}_{3}Li_{3}$, the spins of the last unpaired proton ($j=1/2-$) and the last unpaired neutron ($j=3/2-$) combine to give a $J=1+$ nucleus. In contrast, the spin properties of the two neutrons from the second shell in $^{7}_{3}Li_{4}$ cancel each other out, and the properties of $^{7}_{3}Li_{4}$ are essentially due to the one unpaired $j=3/2-$ proton.

We have previously suggested [11] that the bulk of the energy produced by the E-Cat may be a consequence of Lithium reactions. Here, we hypothesize that the energy is a consequence of an interaction between $^{7}_{3}Li_{4}$ and a proton, resulting in the formation of $^{8}_{4}Be_{4}$, which immediately breaks down into 2 alpha particles. The alpha particles are released with significant kinetic energy, but without gamma radiation. We must reiterate that the energetics of this reaction are still uncertain. On the one hand, we know that there is a strong relative depletion in $^7_3Li_4$, and many of the classical LENR systems utilize Lithium in the electrolyte and produce $^4_2He_2$ particles. On the other hand, alphas were not measured in the Lugano test, while gamma radiation was entirely absent. What therefore can be said about the structure of $^7_3Li_4$, in particular, in relation to the hypothesized: $^7_3Li_4 + p \rightarrow ^8_4Be_4 \rightarrow 2$ alpha reaction?

The lattice structure for the $^7_3Li_4$ ground-state is shown in Figure 4 (left), but this turns out not to be the basis for an explanation of the $^7_3Li_4 + p \rightarrow 2$ alphas reaction. As illustrated in Figure 4 (right), there are four strongly-bonded proton sites on the surface of the ground-state $^7_3Li_4$ (all of which are candidate structures for excited-states of $^8_4Be_4$), but binding of a proton at any of those four sites does *not* lead to a $^8_4Be_4$ geometrical structure containing two alpha tetrahedrons. As a consequence, if a proton were added to the ground-state $^7_3Li_4$ shown in Figure 4, the newly-formed $^8_4Be_4$ isotope would require reconfiguration of nucleon positions and the inevitable release of gamma radiation prior to alpha release. Significant (in excess of 1.0 MeV) gamma radiation has not been observed in the E-Cat, indicating that the ground-state of $^7_3Li_4$ (Figure 4) is an *unlikely* starting point for the relevant reaction. Does the lattice IPM provide no insight?

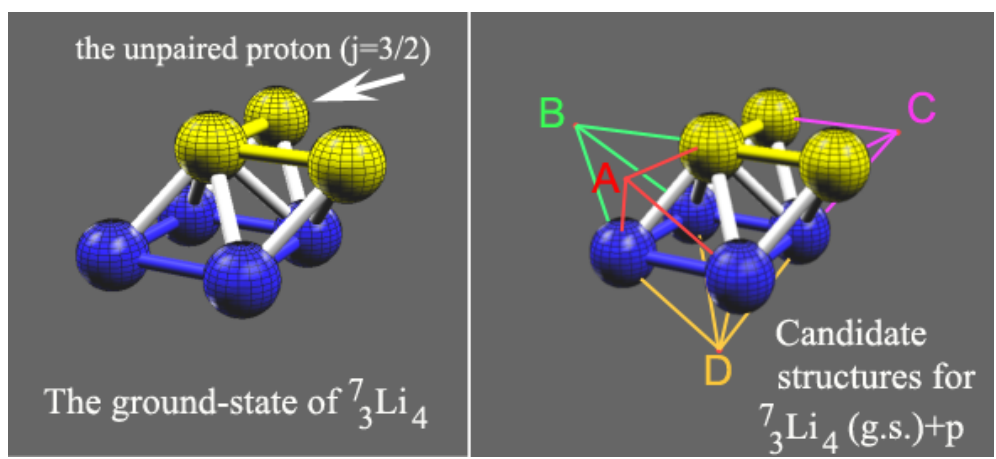

Figure 4: (Left) The ground-state of $^7_3Li_4$. Among several dozen theoretical possibilities, the lattice locations of the last two paired-neutrons and the last unpaired-proton, as shown here, provide a *Jπ* value (3/2-) that is in agreement with experiment. A mirror-image isomeric state has the same properties. (Right) The ground-state of $^7_3Li_4$ and the four lattice locations to which a proton can be added [(A) 1, -3, 1; (B) -3, -3, 1; (C) -1, 3, 1; and (D) -1, -1, -3]. All four produce compact structures (with 3 or 4 nearest neighbor bonds to the $^7_3Li_4$ core), but none produces a $^8_4Be_4$ isotope with two distinct, pre-formed alpha tetrahedrons.

On the contrary, the lattice IPM provides clues when excited-state configurations are considered. There is an unusually low-lying excited state of $^7_3Li_4$ at 0.477 MeV (*Jπ* =1/2-). A $^7_3Li_4$ isotope with those properties can be constructed in the lattice IPM, if the third proton of $^7_3Li_4$ is located at the lower level of protons (lattice coordinates: -1, -1, -3) (Figure 5A). When a fourth proton is added at a neighboring lower proton level (lattice coordinates: -3, -3, -3) (Figure 5B), the newly-formed $^8_4Be_4$ isotope will have a *Jπ* value of 2+, and will contain two distinct alpha tetrahedrons (Figure 5C). As is experimentally known, the first excited-state of $^8_4Be_4$ has *Jπ*=2+ and decays to 2 alpha particles without gamma irradiation (Figure 5D).

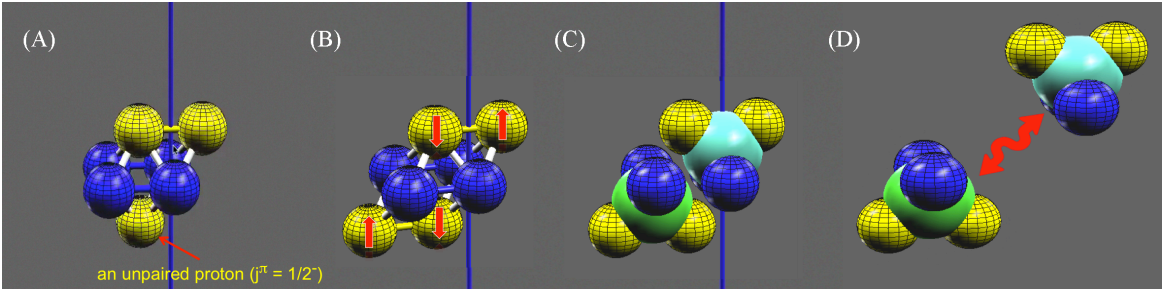

Figure 5: The lowest-lying excited-state of $^7_3Li_4$ (A) has a lattice structure to which an additional proton will produce a two-tetrahedron structure, giving $^8_4Be_4$ (B). The double alpha lattice structure (C) can then break into independent two alpha particles (D), which are released with 17 MeV of angular momentum, but without gamma radiation.

What is of particular interest with regard to the structure shown in Figure 5B is that the $^8Be^4$ configuration is formed from $^7_3Li_4$, where there is one unpaired, spin-up $j$=1/2- proton. By adding one spin-down $j$=5/2- proton to form $^8_4Be_4$, the properties of the two unpaired protons sum to a $J\pi$=2+ state. The $J\pi$=2+ $^8_4Be_4$ isotope is relevant because there are three distinct $^8_4Be_4$ states (a ground-state, $J\pi$=0+, and two $J\pi$=2+ excited states) – all three of which decay to 2 alpha particles without gamma radiation. In other words, unlike the gaseous-phase version of the IPM (where nuclear substructure is essentially absent), the lattice IPM predicts the generation of two alpha particles, unaccompanied by gamma radiation, directly from the $J$=1/2- first excited-state (0.478 MeV) of $^7_3Li_4$.

The two stable Lithium isotopes illustrate the fact that the excited-states of most isotopes arise at energies greater than 2 MeV. The lowest-lying excited state of $^6_3Li_3$ ($J\pi$=1+) is at 2.186 MeV, whereas that for $^7_3Li_4$ ($J\pi$=3/2-) is moderately low at 0.478 MeV. We suggest that it is the presence of odd-A isotopes that makes them more susceptible to configurational changes, in general, and proton accretion, in particular. It must be stated that an energetic justification of the $^7_3Li_4$ + p → $^8_4Be_4$ reaction is still lacking, but if such a reaction is possible, then the geometry of these nuclear states becomes relevant.

### 3.2 Nickel Transmutations

With regard to the transmutations of Nickel, the most obvious reaction mechanisms in NiH systems are listed in Eqs. (11-15). They all entail the addition of one proton to stable Nickel isotopes. If all five reactions actually occur, the net effect would be several β+ decays, and small deposits of stable isotopes: $^{59}_{27}Co_{32}$, $^{63}_{29}Cu_{34}$, and $^{65}_{29}Cu_{36}$ in the E-Cat "ash". In the recent experimental reports [3, 4], significant accumulations of Cobalt and Copper isotopes were **not** found, indicating that reactions (11), (14) and (15) did not occur and therefore that all Nickel isotopes were **not** equally susceptible to transmutation.

$^{58}_{28}Ni_{30}$ (0+)+p → $^{59}_{29}Cu_{30}$ (3/2-)(EC + β+) → $^{59}_{28}Ni_{31}$ (3/2-)(EC+β+) → $^{59}_{27}Co_{32}$ (7/2-)(stable)   (Eq. 11)

$^{60}_{28}Ni_{32}$ (0+) + p → $^{61}_{29}Cu_{32}$ (3/2-) (EC + β+) → $^{61}_{28}Ni_{33}$ (3/2-)(stable)   (Eq. 12)

$^{61}_{28}Ni_{33}$ (3/2-) + p → $^{62}_{29}Cu_{33}$ (1+) (EC + β+) → $^{62}_{28}Ni_{34}$ (0+)(stable)   (Eq. 13)

$^{62}_{28}Ni_{34}$ (0+) + p → $^{63}_{29}Cu_{34}$ (3/2-)(stable)   (Eq. 14)

$^{64}_{28}Ni_{36}$ (0+) + p → $^{65}_{29}Cu_{36}$ (3/2-)(stable)   (Eq. 15)

Moreover, in spite of the fact that the absorption of an alpha particle by $^{58}_{28}Ni_{30}$ would lead directly to an increase in $^{62}_{28}Ni_{34}$ ($^{58}_{28}Ni_{30}$ + α → $^{62}_{30}Zn_{32}$ → $^{62}_{29}Cu_{33}$ → $^{62}_{28}Ni_{34}$), the absence of stable isotopes $^{62}_{30}Zn_{34}$, $^{64}_{30}Zn_{36}$, and $^{66}_{30}Zn_{38}$ in the post-reaction ash indicates that alpha particles (released from the LiH reaction) were **not** absorbed by $^{60}_{28}Ni_{32}$ and

$^{62}_{28}Ni_{34}$. The dramatic increase in $^{62}_{28}Ni_{34}$ must, therefore, be explained through a different mechanism, without implying transmutations for which there is no empirical evidence.

Again, neglecting details of the energetic mechanisms, the main possibility for augmenting $^{62}_{28}Ni_{34}$ abundance is reaction (13). Reaction (13) entails the direct uptake of a proton by $^{61}_{28}Ni_{33}$ (mechanism unknown), leading to $^{62}_{29}Cu_{33}$ (9.7 min), the decay of which would result in the desired isotope, $^{62}_{28}Ni_{34}$. Problematical here is the small abundance of the precursor $^{61}_{28}Ni_{33}$, which accounts for only 1.14% of the Nickel isotopes. The overwhelming abundance of $^{62}_{28}Ni_{34}$ in the ash and the virtual absence of other isotopes might nonetheless be explained as a consequence of the sampling method. Because ToF-SIMS analysis was made on milligram samples obtained specifically at regions observed under the scanning electron microscope to have undergone morphological changes, it is possible that the $^{62}_{28}Ni_{34}$ isotopes recoiled toward the surface of the Nickel grains. If the sample itself was not representative of the Nickel remaining in the E-Cat, the large abundance of $^{62}_{28}Ni_{34}$ would indicate only the participation of $^{61}_{28}Ni_{33}$ in the reaction and its migration to sites that were sampled for isotopic analysis. Further experimental study is needed to clarify the situation.

At the temperature of operation of the E-Cat used in the Lugano test, the Lithium contained in the LiAlH$_4$ is vaporized, and consequently was distributed evenly within the volume of the E-Cat. In contrast, the Nickel fuel remained in a solid or liquid state. At the time of sampling after one month of operation, Nickel was found to be encrusted on the internal surface of the reactor, from which a 2 mg sample of "ash" was obtained near to the center of the charge. Starting with an initial charge of approximately 1 gram, it cannot be said that the 2 mg sample was necessarily representative of the entire Nickel charge, but it remains to be explained how the isotopic ratios in the 2 mg sample show predominantly $^{62}_{28}Ni_{34}$.

Isotopes with extremely low-lying excited states are of particular interest in LENR research because they exhibit quantal transitions from one nucleon state to another with minimal external input. In this regard, the lowest-lying excited-state of one of the most stable isotopes in the Periodic Table of Elements, $^{61}_{28}Ni_{33}$ (a *J*=5/2+ state at 0.0674 MeV), is a likely candidate for energy release in response to low-level thermal agitation. That excitation energy stands in contrast to all of the stable even-even isotopes of Nickel whose lowest-lying excited-states are typically 20~40 fold higher (>1.3 MeV).

There is, in fact, a small number of comparable excited states in stable isotopes across the Periodic Table, notably, $^{103}_{45}Rh_{58}$ (*J*=7/2, 0.0397 MeV) and $^{105}_{46}Pd_{59}$ (*J*=3/2, 0.280 MeV), both of which have been implicated in prior LENR research. Noteworthy, however, is the fact that their natural abundances are extremely low. Specifically, $^{103}_{45}Rh_{58}$ is present in the Earth's crust at a level of 0.0010 mg/kg, and $^{105}_{46}Pd_{59}$ at 0.0033 mg/kg, whereas $^{61}_{28}Ni_{33}$ is present at a level of 0.9576 mg/kg. These relative abundances mean that technological application of their LENR capacities would be, respectively, 1000 times and 300 times more expensive for Rhodium and Palladium relative to Nickel.

As noted above, nuclear reactions involving low-lying excited-states are speculative insofar as the initiating "cold fusion" reaction demands the accretion of a proton by a stable nucleus at temperatures not normally reached except in "hot fusion" conditions. The question arises whether or not an energetically favorable mechanism might initiate MeV nuclear events. In this context, the relatively low-energy $^{7}_{3}Li_{4}$ + p reaction, leading to 17 MeV alpha release, is of considerable interest.

# 4 Discussion

The Nickel-LiAlH$_4$ system known as the E-Cat is one of several dozen LENR configurations for which excess heat has been experimentally demonstrated [1, 2]. The E-Cat is, however, apparently unique in allowing for the reliable production of significant energy using relatively inexpensive materials. Although its main source of energy appears to be the $^7_3Li_4(p, \alpha)\alpha$ reaction, the recently reported transmutations [3] are strongly suggestive of two distinct types of LENR – neither of which is easily explained in traditional nuclear physics. Specifically, both of the most likely reactions induced in the E-Cat entail nucleon uptake by stable, odd-A isotopes. The coincidence that both $^7_3Li_4$ and $^{61}_{28}Ni_{33}$ are stable $J$=3/2- isotopes with low-lying excited states (<0.5 MeV) is suggestive that the unanticipated phenomena of LENR may be a consequence of the detailed substructure of easily-excited stable isotopes. Particularly in light of the fact that the quantal states of nucleons in the IPM have a straightforward lattice geometry [5], from which nuclear $J\pi$-values and magnetic moments can be predicted, we conclude that it is worthwhile to examine the largely-overlooked nuclear structure aspects of LENR. Stated conversely, as important as the solid-state environment and the surrounding electromagnetic field is for inducing nuclear effects, the nuclear reactions themselves appear to occur only in a few specific isotopes and involve only a few specific quantal transitions. If the excitation of stable nuclei to low-lying excited-states is indeed an essential prerequisite of LENR phenomena, it would not be surprising that LENR effects can occur in very different solid-state/chemical environments, provided only that the necessary proton/deuteron constituents can be brought into contact with the unusually-reactive low-lying excited-states of substrate nuclei.

### 4.1 A Plethora of Cold Fusion Theories

Many quantum-theory-based hypotheses have been advocated to explain cold fusion phenomena. Gullstroem [12] has proposed a neutron exchange mechanism to explain specifically the E-Cat transmutation effects. Muelenberg [13] and Muelenberg and Sinha [14] have proposed a "lochon" (local charged boson) model as a means for overcoming the Coulomb repulsion between protons, deuterons and other nuclei. Previously, Ikegami and others [15-21] have proposed that alpha particles can be generated by Lithium (following proton accretion or deuteron stripping). Quantitative results and a consensus concerning their significance in specific experimental contexts are yet to be obtained, but such theoretical work will eventually be of fundamental importance in order to provide an energetic justification for LENR phenomena.

### 4.2 A Common Theme

What all LENRs have in common are unanticipated nuclear events that traditional nuclear physicists would categorically maintain to be impossible. There is indeed little doubt that the "central dogma" of atomic physics:

Neutrons ←→ Protons → Electrons

generally holds true. Nuclei have strong influence on extra-nuclear events, but not vice versa – primarily because electron transitions occur at the level of several electron-Volts (eV), while nuclear transitions typically occur at the level of millions of electron Volts (MeV). However, LENR phenomena, in general, and the recently reported transmutation results [3], in particular, clearly indicate that there are circumstances where nuclear reactions can be initiated in chemical systems at relatively low-energies.

It is noteworthy, moreover, that the well-known Mössbauer effect also entails "violation" of the central dogma, but is today an established part of nuclear physics. As Wertheim noted in 1960 [22]:

> "Nuclear physicists have a strong and understandable tendency to ignore the chemical binding of the atoms whose nuclei they investigate. This is based on the fundamentally sound precept that the energies involved in nuclear reactions are so much larger than the energies of chemical binding that the atom may well be thought of as a free atom when analyzing nuclear events." (p. 1)

This "precept" was, however, found to be violated in the Mössbauer effect and immediately led to a suitable expansion of the central dogma of atomic physics to include a small set of low-energy solid-state phenomena in which electron effects can influence nuclear effects:

Neutrons ←→ Protons ←→ Electrons

The phenomena of late 20th / early 21st century "cold fusion" physics (LENR) appear to take place in a similar energetic context.

Be that as it may, the changes in natural isotopic abundances in the E-Cat and other "cold fusion" systems are unambiguous indication that nuclear reactions have occurred – reactions that require explanation that is consistent with nuclear structure theory. Clarification of precise mechanisms will undoubtedly require measurements of low-level gamma radiation within LENR systems to establish unambiguously which quantum states of which nucleons in which isotopes are involved.